\documentclass[doublecol]{epl2} 

\usepackage{epsfig}
\usepackage{citesort}

\newcommand{\eq}{\begin{equation}}
\newcommand{\eqx}{\end{equation}}
\newcommand{\eqn}{\begin{eqnarray}}
\newcommand{\eqnx}{\end{eqnarray}}

\title{Atomic clusters of various sizes irradiated with short intense pulses of VUV radiation}
\shorttitle{Size dependent dynamics within VUV irradiated clusters} 

\author{B. Ziaja \inst{1,2} \and  H. Wabnitz \inst{1} \and E. Weckert \inst{1}
\and  T. M\"oller \inst{3}}
\shortauthor{B. Ziaja \etal}

\institute{                    
  \inst{1} Hamburger Synchrotronstrahlungslabor, Deutsches
  Elektronen-Synchrotron - Notkestr. 85, D-22603 Hamburg, Germany,
  e-mail: ziaja@mail.desy.de
  
  \inst{2} Department of Theoretical Physics, Institute of Nuclear Physics -
  Radzikowskiego 152, 31-342 Cracow, Poland
  
  \inst{3} Technische Universit\"at Berlin, Institut f\"ur Atomare Physik und 
  Fachdidaktik - 10623 Berlin, Hardenbergstrasse 36, Germany
}
\pacs{41.60.Cr}{Free-electron lasers}
\pacs{52.50.Jm}{Plasma production and heating by laser beams}
\pacs{52.30.-q}{Plasma dynamics and flow}
\pacs{52.65.-y}{Plasma simulation}

\abstract{Non-equilibrium processes following the irradiation of atomic 
clusters with short pulses of vacuum ultraviolet radiation are modelled using kinetic Boltzmann equations. The dependence of the ionization dynamics on the cluster size is investigated. The predictions on: (i) the maximal and average ion charge created, (ii) ion charge state distribution, (iii) average energy absorbed per atom, (iv) spatial charge distribution, and (v) thermalization scales are obtained for spherical xenon clusters  containing: $20$, $70$, $2500$ and $90000$ atoms. These clusters were exposed to single rectangular pulses of vacuum ultraviolet radiation of various pulse intensities, $I\sim 10^{12}-10^{14}$ W/cm$^2$ and durations, $\Delta t \leq 50$ fs, at a fixed integrated radiation flux of $F\equiv I\cdot\ \Delta t=0.4$ J/cm$^2$. The results obtained are found to be in good agreement with the available experimental data, especially the dependence on the cluster size, if it is assumed that the ions from the positively charged outer layer of the cluster constitute the dominant contribution to the experimentally measured ion charge state distribution.}

\begin{document}

\maketitle

Atomic clusters are physical objects of nanometer size, consisting of a few up to tens of thousands of closely packed atoms. Their physical properties put them onto the border between solid state and gas phase. Clusters are excellent objects to test the dynamics within samples irradiated with short wavelength radiation from free-electron-lasers (FELs) \cite{tesla,desy2006,slac,jap}. By studying the size dependence the influence of the atomic and condensed matter effects on the ionization dynamics can be explored. Cluster studies are important for planned experiments with FELs within the solid state physics, material sciences and for studies of the extreme states of matter \cite{xfelinfo2007}. Accurate predictions on the ionization, thermalization and expansion timescales within irradiated samples that can be obtained with cluster experiments are needed for exploring the limits for experiments on single particle diffraction imaging \cite{l1,miao,gyula1,plasma4,liver1en,chapman}.

In ref.~\cite{ziajab2} we investigated the ionization dynamics within  $Xe_{2500}$ clusters irradiated with rectangular vacuum ultraviolet (VUV)
pulses of five different integrated pulse energy fluxes: $F=0.05, 0.3, 0.84,1.25,1.5$ J/cm$^2$. Pulse intensities were, $I \sim 10^{12}-10^{14}$, W/cm$^2$ and pulse duration was $\leq 50$ fs. For each flux the intensities and pulse durations were chosen in order to match the condition: $I \cdot \Delta t=F$. Various intensities were tested at each fixed integrated radiation flux. The predictions obtained were then averaged over the number of events. This procedure enabled us to account for the non-linear response of the system to various pulse lengths and pulse intensities at higher radiation fluxes. This scheme followed the experimental analysis: the experimental data were obtained after averaging the single shot data obtained with FEL pulses of a fixed integrated radiation flux. 

Here we study the non-equilibrium ionization dynamics within clusters of various sizes ranging from 20 to 90000 atoms. These clusters are exposed to single rectangular VUV pulses of wavelength $\sim 100$ nm
(photon energy $\sim 13$ eV) and duration $\leq 50$ fs. Various pulse intensities are tested, however, the integrated radiation flux is kept constant, $F \equiv I\cdot \Delta t=0.4$ J/cm$^2$. For comparison, if the pulse length is $\Delta t=50$ fs, this integrated radiation flux can be achieved 
with the pulse intensity, $I=8\cdot 10^{12}$ W/cm$^2$. The above pulse parameters correspond to  those of the cluster experiment performed at the FLASH facility at DESY \cite{desy,desy2,desy3}. Results of this experiment show that ionization dynamics within the irradiated clusters strongly depends on cluster size. 

The present study applies the theoretical framework defined in ref.~\cite{ziajab2}. It completes the analysis performed in ref.~\cite{ziajab2} by evaluating the effect of the cluster size on the ionization dynamics.

Our simulation tool solves the  semiclassical kinetic equations for electron and ion densities within the irradiated samples. Charges (represented as charge densities) interact with the mean electromagnetic field created by all charges and also with the laser field. Microscopic interactions enter these equations as rates. These rates are included into the two-body collision terms, and are estimated either from the experimental data or with theoretical models.
Our model treats the following interactions: photoionizations, collisional ionizations, elastic scatterings of electrons on ions, inverse bremsstrahlung
(IB) heating, shifts of energy levels within atomic potentials due to the plasma environment, and the shielded electron-electron interactions. 

This kinetic approach has two important features: (i) in contrast to hydrodynamic models, it can also follow the non-equilibrium sample evolution. Therefore it can deliver information on the timescales needed for reaching the local thermodynamic equilibrium (LTE) within a system, and (ii) it is computationally efficient for large samples. Further details of the method are discussed in refs.~\cite{ziajab,ziajab2}. Based on the conclusions of 
ref.~\cite{ziajab2}, we model the IB process with the enhanced IB rate and assume the plasma shifted energy levels of the atomic potentials. 
 
Below we discuss the results of our simulations performed for small and large clusters. The evolution of a sample hit by a single VUV pulse can be separated into two phases. The non-equilibrium ionization phase  starts after the sample is exposed to the laser radiation and lasts until the saturation of ionizations from ground states is reached. Its duration may last up to several tens of femtoseconds depending on cluster size, pulse length (assumed to be $\leq 50$ fs) and pulse intensity (assumed to be $\sim 10^{12}-10^{14}$ W/cm$^2$). The semi-equilibrium expansion phase that follows after the ionization phase is much longer. Its timescale is of the order of picoseconds. Here we follow only the ionization phase and stop the simulation at entering the expansion regime.

During the ionization phase electrons are released from atoms and ions. While they stay inside the sample, they are efficiently heated with the inverse bremsstrahlung process \cite{santra,santra1}. Energetic electrons can then leave the cluster, creating an attractive Coulomb potential. Slower electrons are kept inside the sample, and the Coulomb attraction forces move these electrons towards the centre of the cluster. Therefore, at the end of ionization phase the charge distribution within the sample shows a characteristic layer structure consisting of a neutral core and of a positively charged outer shell. This inhomogeneous charge structure was also described in 
refs.~\cite{gyula,plasma,siedschlag,rost1,ziajab2}. In our case we observe that the cluster core is dominated by ions of the highest charges, however, the net charge of the core remains equal to zero. The core neutrality 
is due to the presence of quasi-free electrons bound within the core. The positively charged surface layer consists of ions of various charges, in particular, it contains ions of lower charges. 
 
\begin{figure}
\vspace*{0.5cm}
\centerline{\epsfig{width=7cm, file=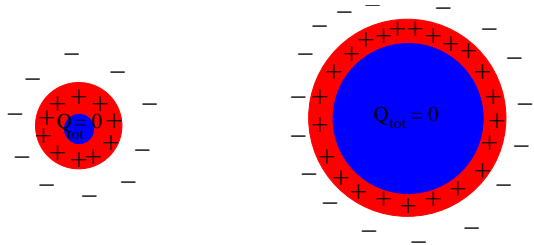}
}
\caption{Schematic plots of charge distribution within an irradiated 
small cluster (left), and an irradiated large cluster (right) 
during the ionization phase. Positively charged outer shell coats the neutral core. Some electrons have escaped from the cluster (outer ionization).}
\label{cluster}
\end{figure}

\begin{figure}
\vspace*{0.5cm}
\centerline{a)\epsfig{width=4cm, file=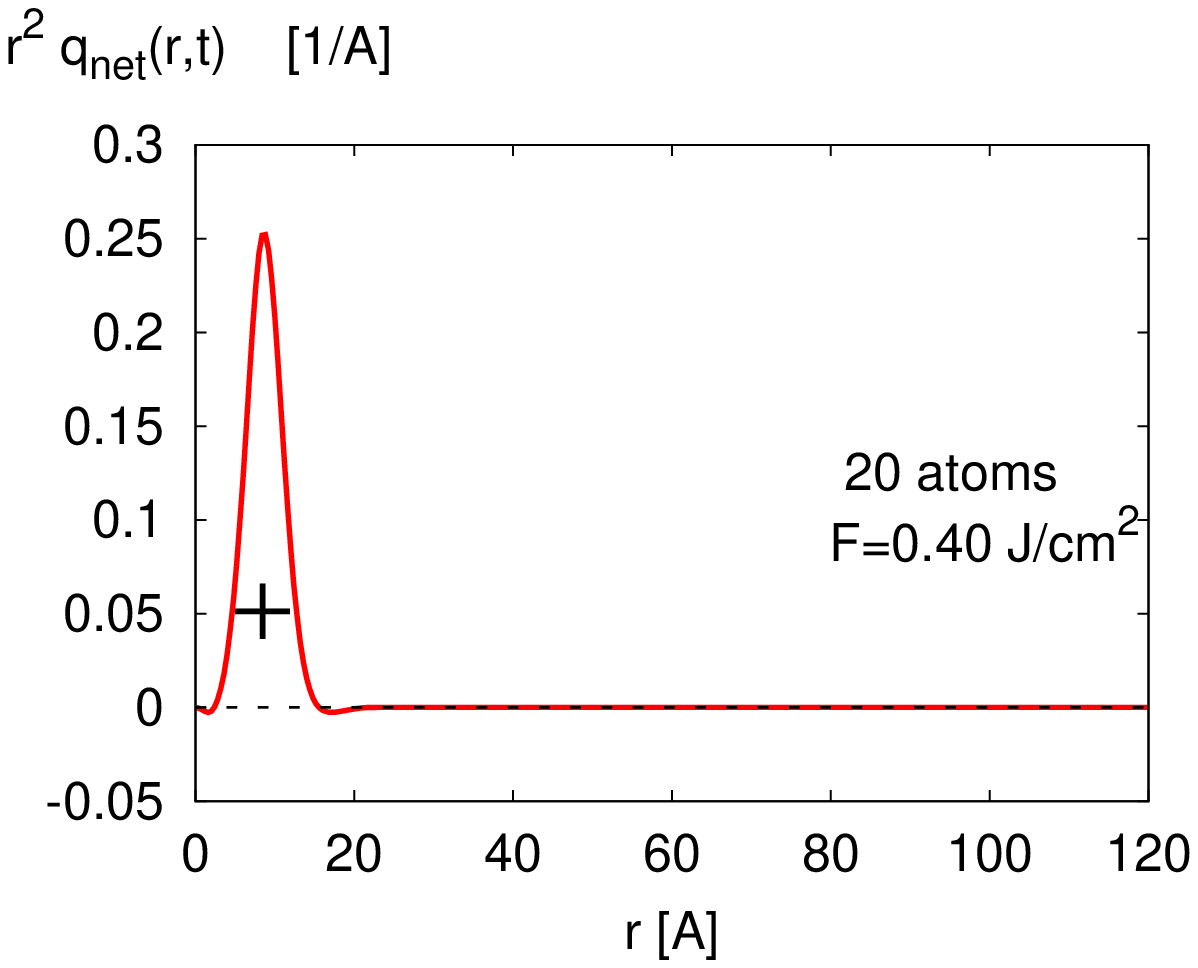}
b)\epsfig{width=4cm, file=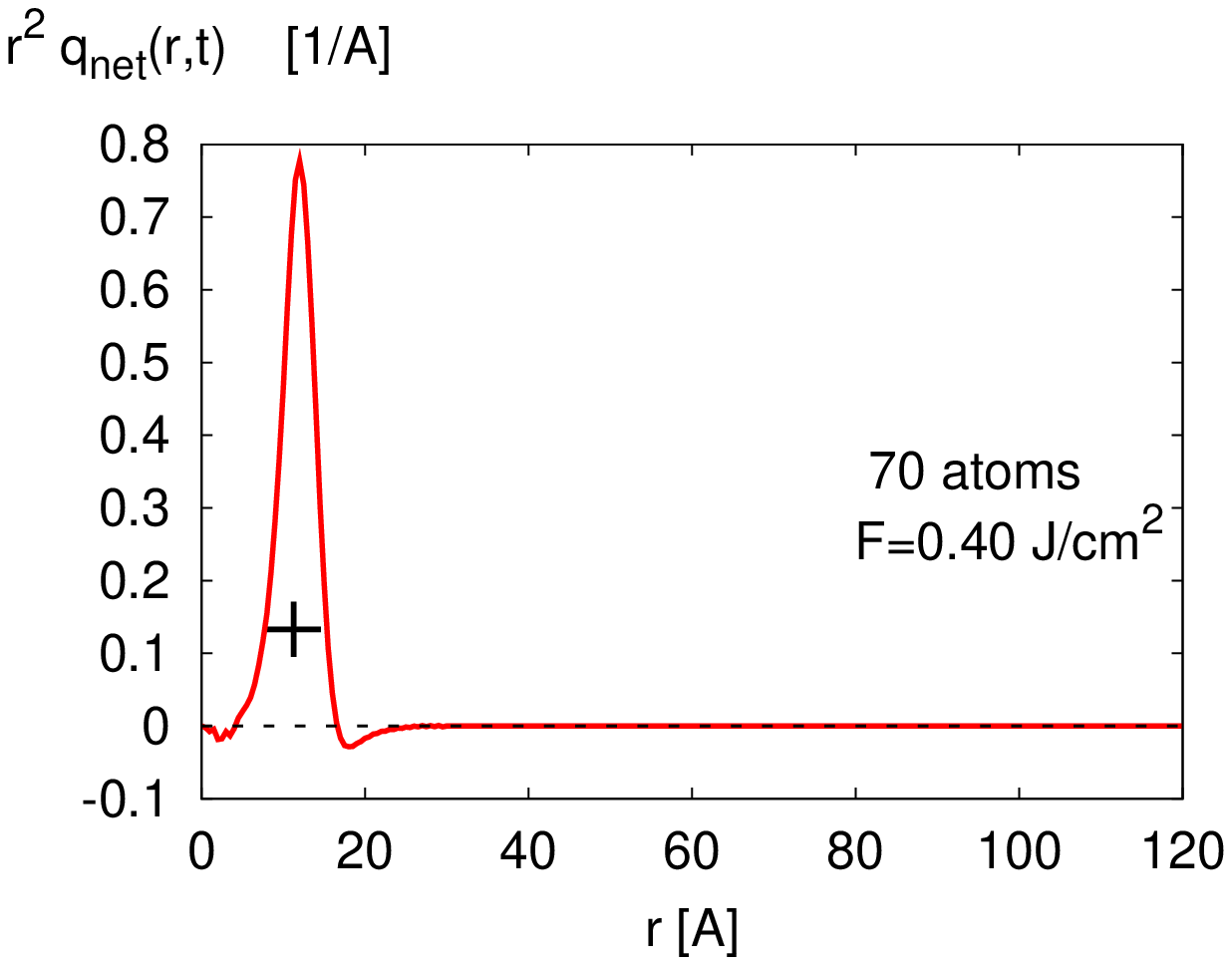} }
\vspace{0.5cm}

\centerline{c)\epsfig{width=4cm, file=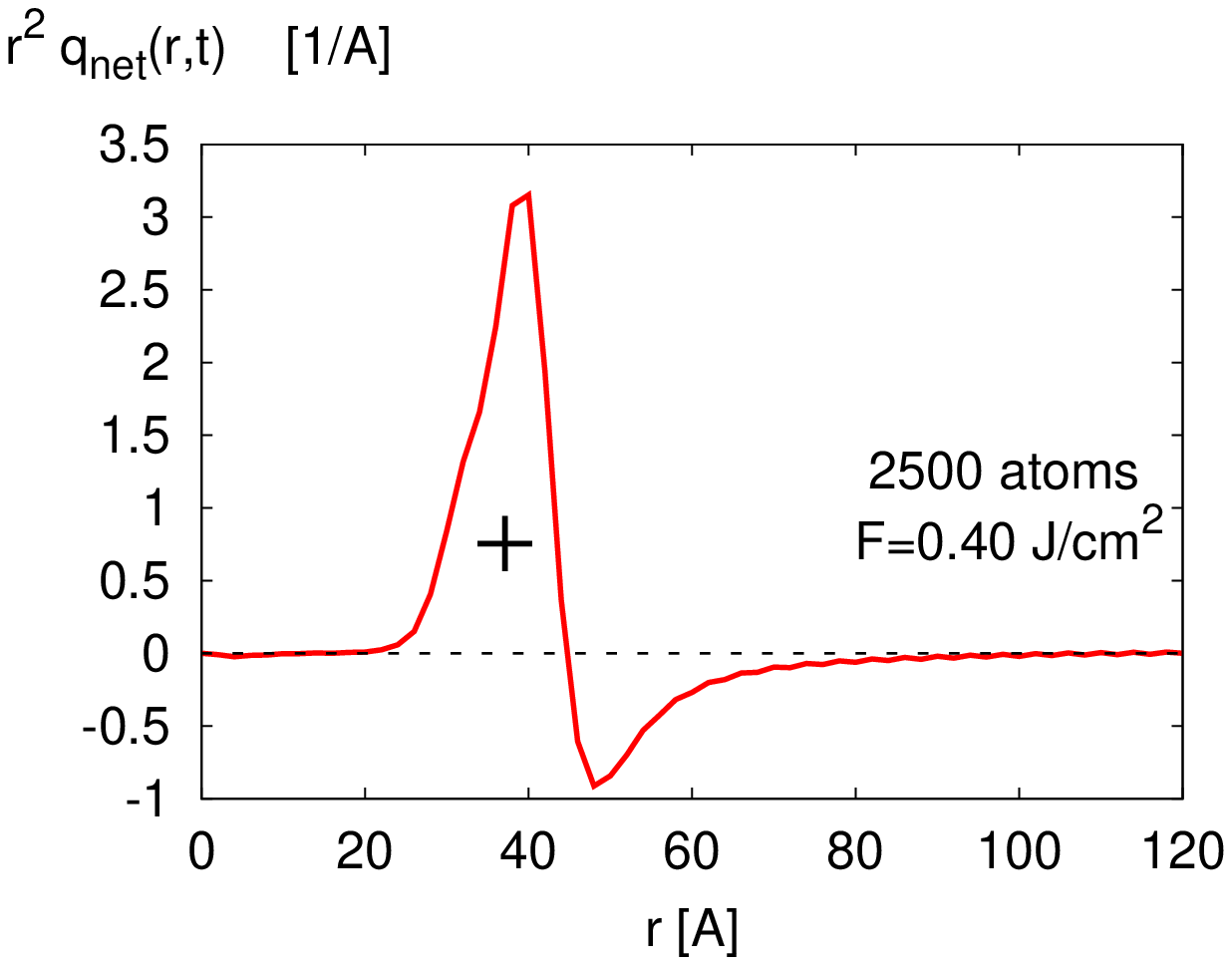}
d)\epsfig{width=4cm, file=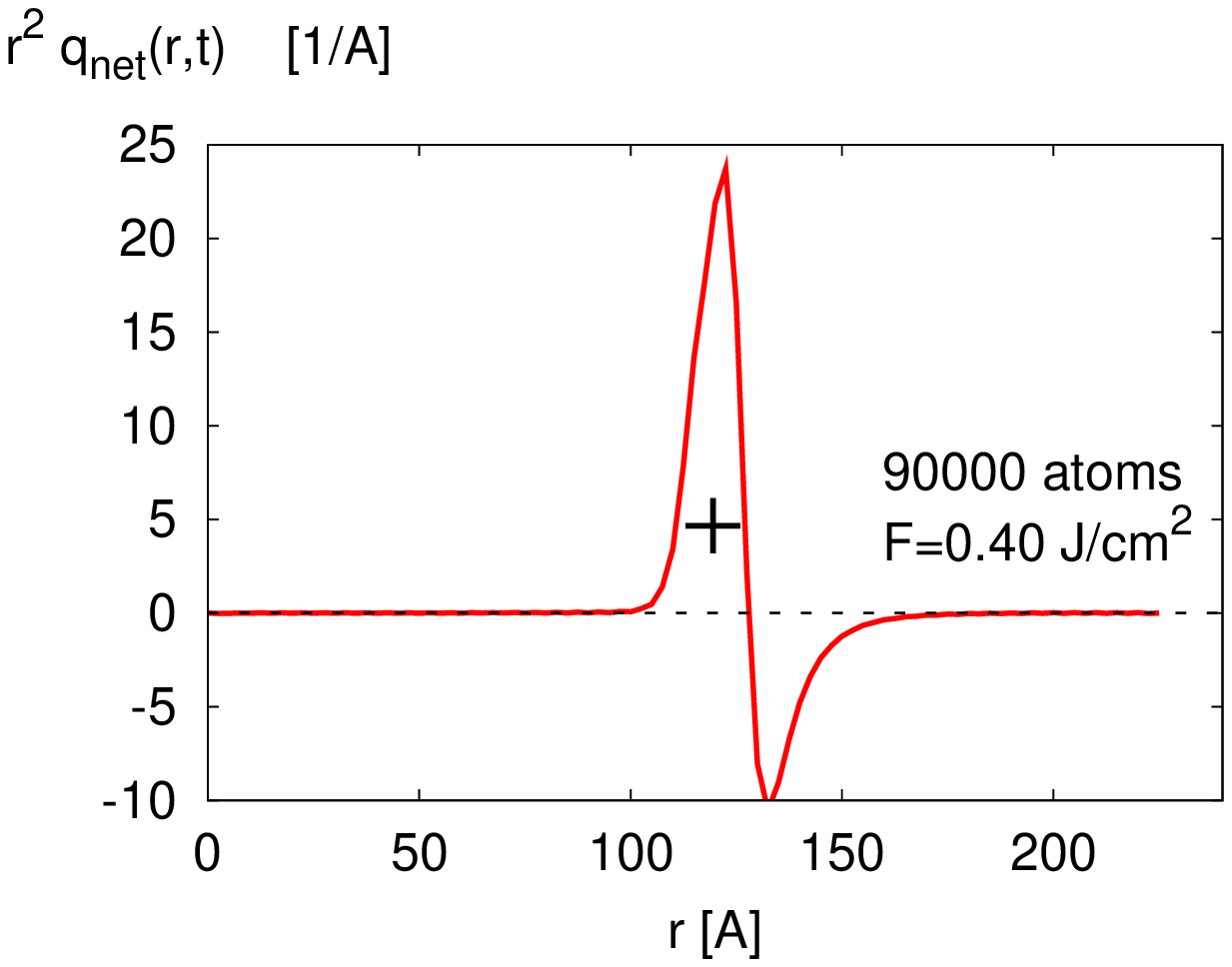} }
\vspace{0.5cm}

\caption{Formation of the outer shell of ions during the ionization phase. The charge density is defined as $q_{net}(r,t)=\sum_{i=1}^{N_J}\, i\cdot n_i(r,t) -n_e(r,t)$, where $n_i(r,t)$ and $n_e(r,t)$ denote ion and electron densities.
The maximal ion charge is, $N_J=8$. Plots show the neutral core and the positively charged outer shell created within clusters of various sizes: a)-b) extensive outer shell for smaller clusters, c)-d) narrow outer shell for large clusters. Some electrons have escaped from the cluster (outer ionization)}
\label{outer}
\end{figure}

We find that within smaller samples a large fraction of the electrons released during ionization processes is able to leave the sample early in the exposure. The width of the positively charged outer shell is then large, comparing to the radius of the neutral core (fig.~\ref{cluster} (left), figs.~\ref{outer}a-b). Therefore, we expect the Coulomb explosion to be the dominating mechanism of the overall cluster expansion.

Within larger samples the number of energetic electrons that are able to leave the cluster during the ionization phase is small with respect to the total number of electrons released. Therefore, the width of the outer ion shell is small in comparison to the radius of the neutral core (fig.~\ref{cluster} (right), figs.~\ref{outer}c-d). It is expected that after the Coulomb-driven escape of the outer ions the core will slowly expand due to the hydrodynamic pressure of electrons and ions. Recombinations and ionizations (to and from excited states) will still occur within the neutral core. These processes may reduce the total ion charge within the core \cite{ziajab2}.

The Coulomb expansion is faster than the hydrodynamic expansion. The difference in these expansion rates can be useful at planning experiments on single particle imaging \cite{plasma4,liver1en}. If a biomolecule could be coated with a layer of some other element or compound \cite{liver1en,water}, only the outer part of the layer would undergo fast Coulomb explosion. The remaining neutral core containing the particle under investigation would expand hydrodynamically, slower than the outer shell. Compared to the case of an uncoated sample, the coating would increase the radiation tolerance of the sample. This idea was investigated in ref.~\cite{liver1en} for samples irradiated with X-rays. The simulations performed within our study support the above scenario. 

Next we compare our results to the experimental data obtained with FLASH at DESY at $100$ nm radiation wavelength \cite{desy,desy7}. The experimentally estimated histograms of ion charge and predictions on the average energy absorption were obtained from the averaged time-of-flight (TOF) spectra. We remind here that the TOF detector could record charged particles (ions) only. There are no data on neutrals (atoms) available from these measurements. The experimental ion intensities were obtained by integrating and averaging the TOF signal over 100 subsequent FEL pulses. Those intensities were not corrected for the relative geometric acceptances of the TOF detector or the MCP detector efficiencies for different charge states.
 
\begin{figure}
\vspace*{0.5cm}
\centerline{a)\epsfig{width=4cm, file=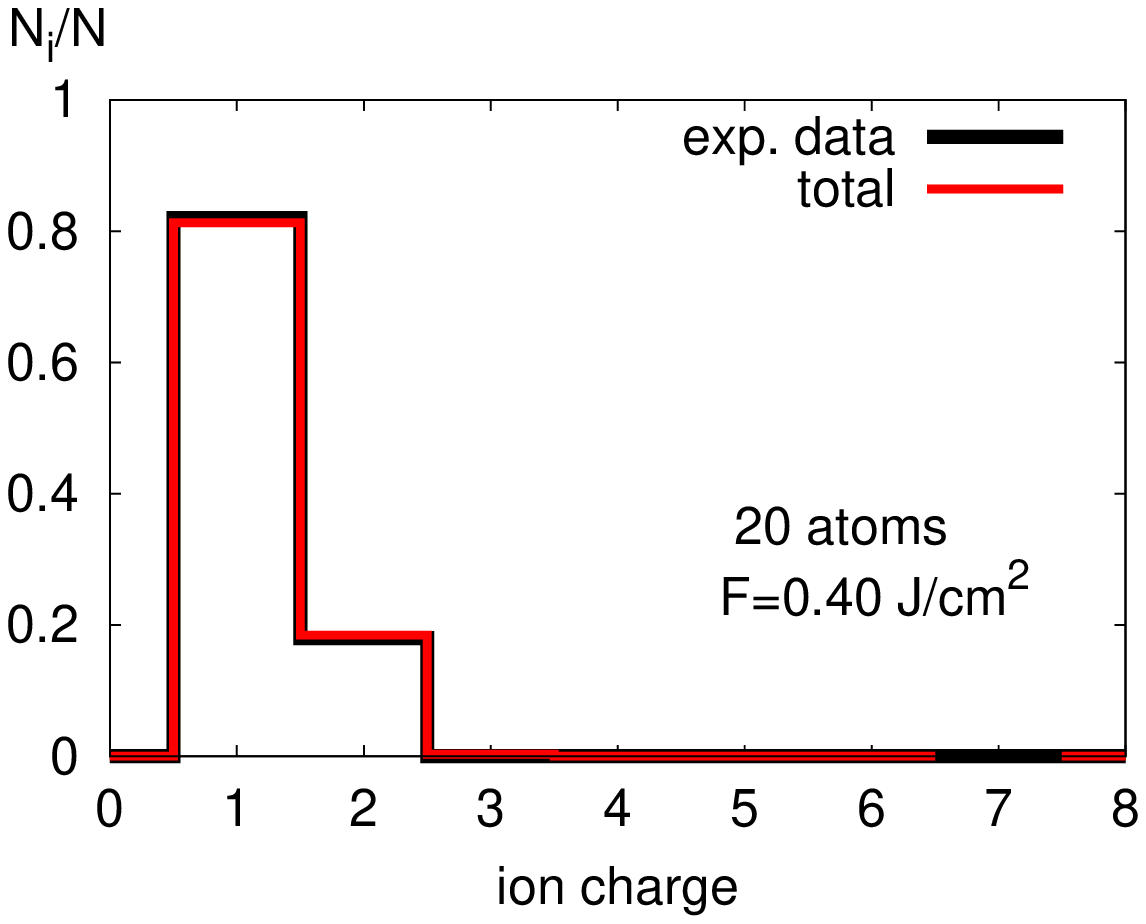}
b)\epsfig{width=4cm, file=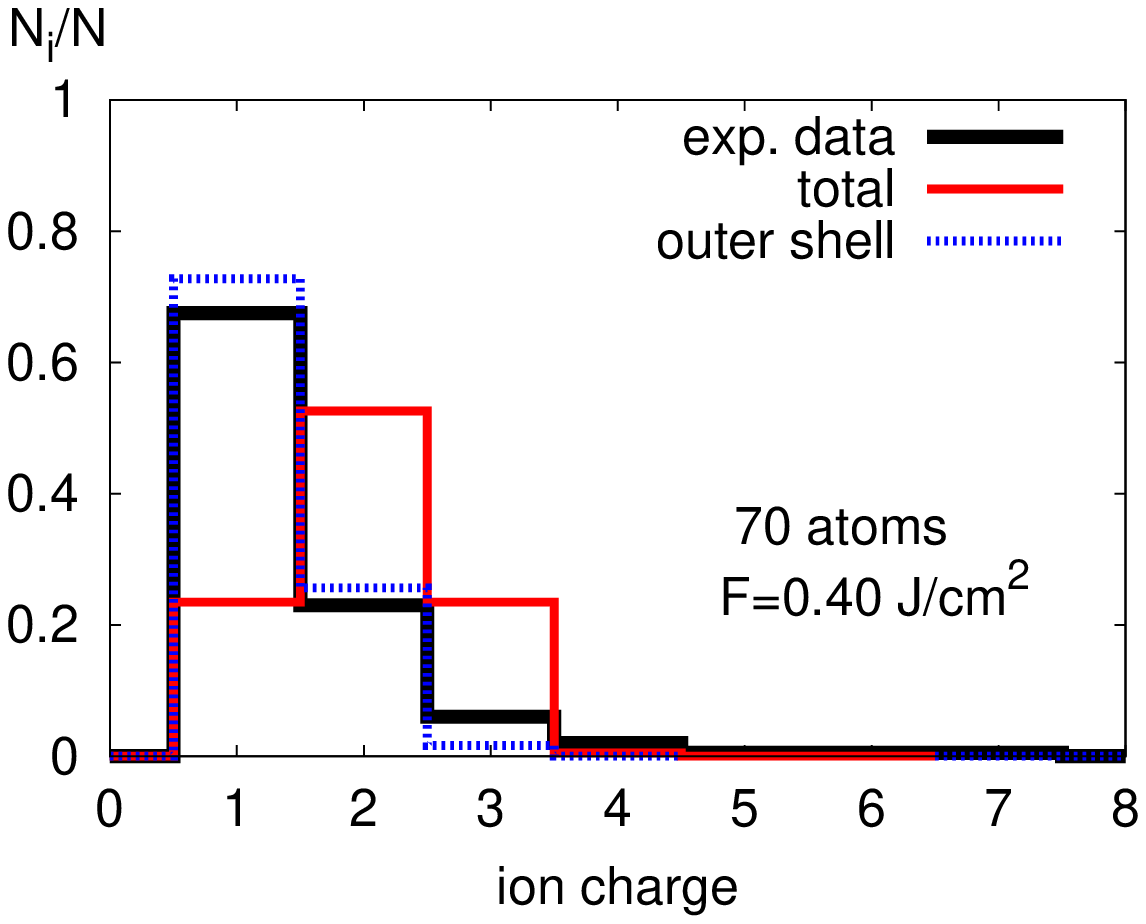} }
\vspace{0.5cm}

\centerline{c)\epsfig{width=4cm, file=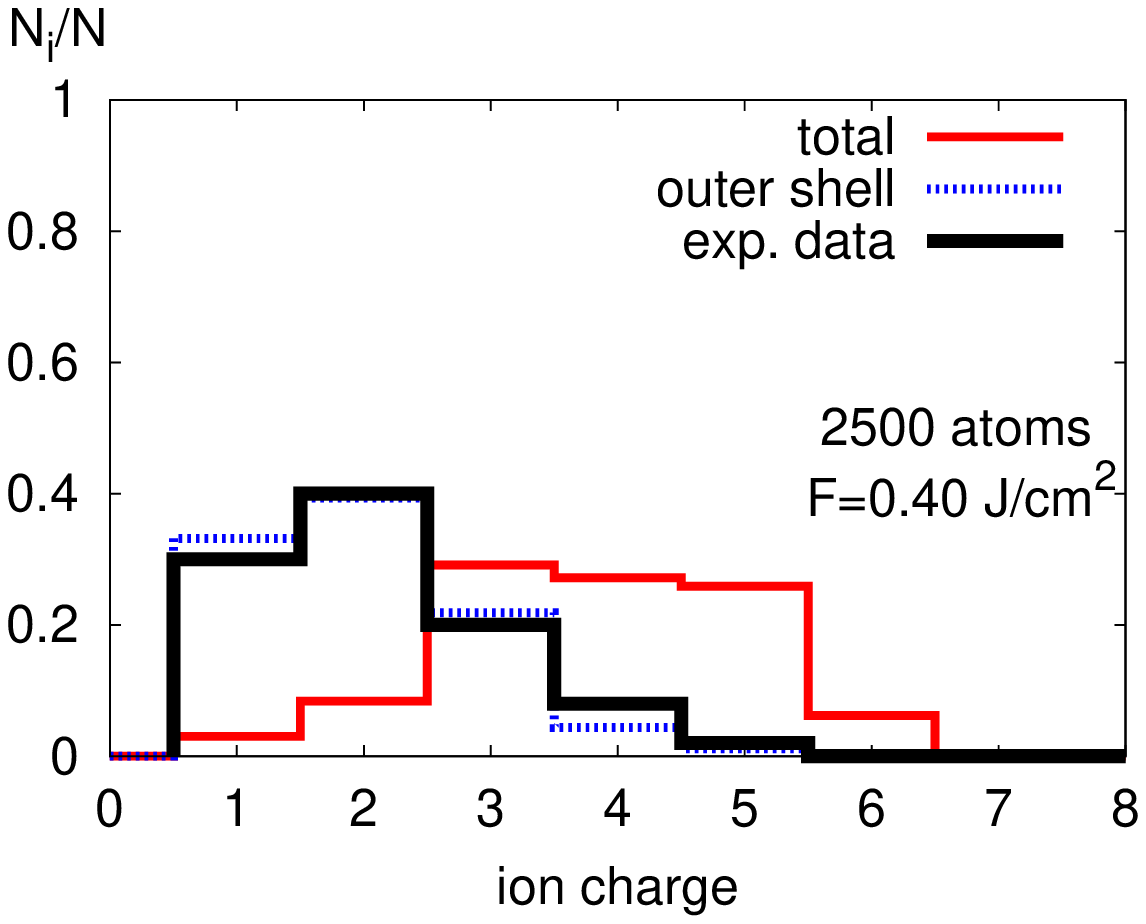}
d)\epsfig{width=4cm, file=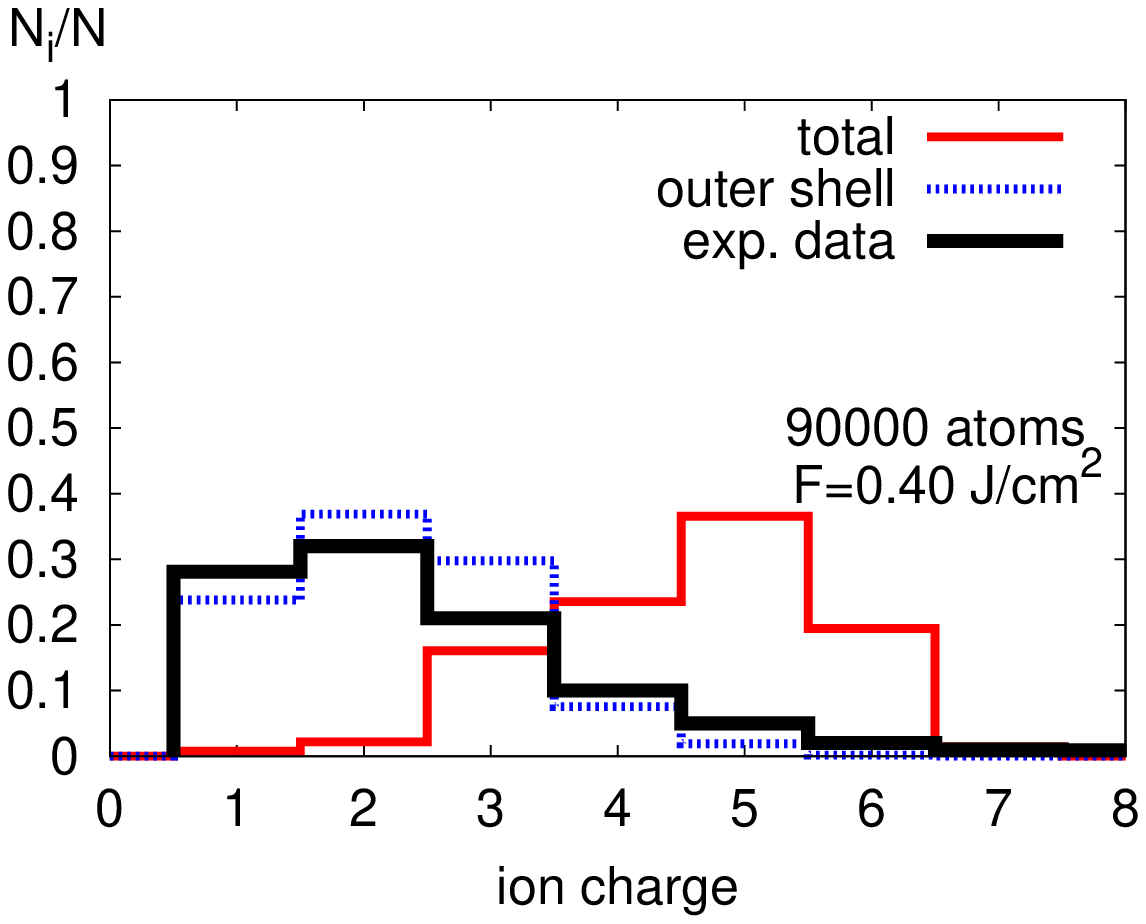} }
\vspace{0.5cm}

\caption{Ion fractions, $N_i/N$, within the irradiated xenon clusters at the end of ionization phase. The number, $N_i$ denotes here the number of $Xe^{+i}$ ions, and the number $N$ is the total number of ions. At $t=0$ fs these clusters contained: a) 20 atoms, b) 70 atoms, c) 2500 atoms and d) 90000 atoms. For $Xe_{20}$ cluster the experimental and theoretical results overlap. For larger clusters b)-d) also the ion fractions within the positively charged outer shell are shown. The experimental results are in agreement with the ion fractions from the positively charged outer shell.}
\label{charge}
\end{figure}

First we present the simulation results for smaller clusters consisting on average of 20 and 70 Xe atoms. As stated previously, a large fraction of the electrons released during the ionization processes can leave the small clusters early in the exposure. The remaining electrons are not heated efficiently via IB processes due to their low density within the cluster. Only charge states up to +2 ($N=20$ atoms) and +4 ($N=70$ atoms) are observed (fig.~\ref{charge}a-b). The predictions obtained are in agreement with the experimental findings. The agreement is better for clusters of 20 atoms, for cluster of 70 atoms we find a discrepancy between the predictions and the data. This discrepancy can be
reduced, if we assume that the ions from the positively charged surface layer
of the cluster give the dominant contribution to the experimentally measured
ion charge spectra. This assumption has been discussed in detail in 
ref.~\cite{ziajab2}. The resulting outer shell predictions are shown in 
fig.~\ref{charge}b). Experimental data are within the model predictions.

Further, we investigate large clusters built of 2500 and 90000 xenon atoms. After the irradiation, only a small fraction of the electrons is able to escape from these large samples. The width of the outer shell created is small with respect to the radius of the neutral core. Following the discussion in 
ref.~\cite{ziajab2}, in fig.~\ref{charge}c-d we plot the ion fractions obtained within the whole cluster and within its outer shell at the end of the ionization phase.  Our predictions on the maximal ion charge are in agreement with experimental predictions, except of the absence of charges +8 within the largest cluster, $N=90000$ atoms. The trend of the size dependence is correct: at fixed pulse energy the maximal ion charge created increases with the cluster size. 

We note that the ion fractions observed within the whole cluster at the end of the ionization phase will not correspond to those recorded by the detector during the experiment. Ions from the positively charged surface layer will be the first ones to escape from the sample, and they will reach the TOF detector with the unchanged charge state distribution. For the cluster core the situation is different. At the end of ionization phase the core is a dense system of quasi-free electrons and ions, where recombination and ionization processes occur. During the long picosecond expansion phase charges within the core have enough time for an efficient recombination. Therefore, the remnants of the core will be weakly charged or neutral. They will reach the detector at the end of the ionization phase, and then modify the ion charge state distribution recorded in the detector by increasing the participation of lower charges. This mechanism should be quantitatively verified with an expansion code, e. g. hydrodynamic code. This is, however, beyond the scope of the present study. 

\begin{figure}
\vspace*{0.5cm}
\centerline{\epsfig{width=7cm, file=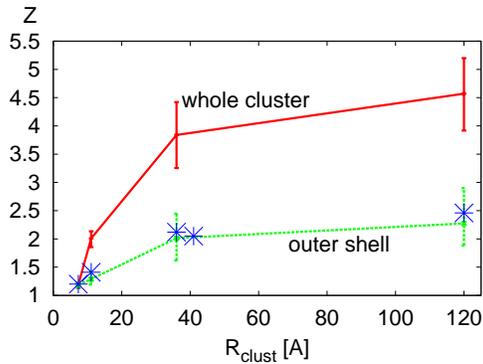}}
\caption{Average charge, $Z$, created within the whole irradiated cluster (red errorbars) and within the outer shell (green errorbars) as a function of the cluster radius, $R$. These estimates were obtained with pulses of different intensities and lengths but of fixed integrated radiation flux, $F=0.4$ J/cm$^2$, and then averaged over the number of pulses. Errorbars estimate maximal errors. Experimental data are plotted with stars.}
\label{lad}
\end{figure}

We focus now on the ion charge state distribution observed within the outer shell for large clusters, $N=2500,90000$ atoms. They are in good agreement with the experimental data (fig.~\ref{charge}c-d). This indicates that the recombination within the cluster core is efficient during the expansion phase. Also, the average charge predicted within the whole cluster and within the outer shell supports this observation. This charge is plotted as a function of cluster radius in fig.~\ref{lad}. The charge calculated within the outer shell follows the experimental results and starts to saturate for clusters of radius $\geq 40$ \AA, i. e. built of more than $3300$ Xe atoms. 

Within small clusters, recombination processes will not be efficient due to the low electron density within the core. Therefore one can expect that the ion fractions recorded at the end of ionization phase within the whole cluster will approach those detected in experiment. Our results on small clusters of $N=20, 70$ atoms support this scenario (fig.~\ref{charge}a,b).

\begin{figure}
\vspace*{0.5cm}
\centerline{\epsfig{width=7cm, file=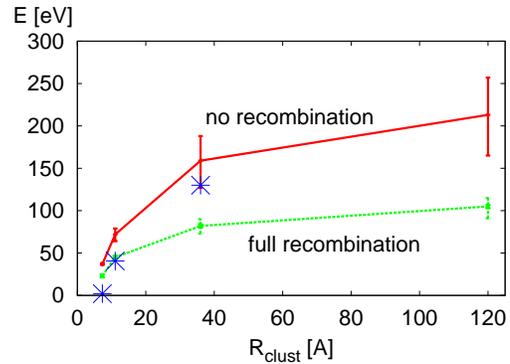}}
\caption{Average energy absorbed per atom, $E$, within the irradiated cluster as a function of the cluster radius, $R$. Upper and lower limits (red and green errorbars) for the absorbed energies are estimated within our model. These estimates were obtained with pulses of different intensities and lengths but of fixed integrated radiation flux, $F=0.4$ J/cm$^2$, and then averaged over the number of pulses. Errorbars estimate maximal errors. Experimental data are plotted with stars.}
\label{energy}
\end{figure}

Below we show also the average energy absorbed per atom estimated with our model as a function of cluster radius (fig.~\ref{energy}). With our model we can only obtain the upper and the lower limit for the absorbed energy. The upper limit assumes that during the further expansion of the sample no recombination processes are occurring. The lower limit gives an estimate of the energy absorbed per atom in case of the full neutralization of the sample during the expansion (full recombination). These limits are compared to the available experimental data on the average ejection energy per atom. Experimental data for larger clusters are within the model estimates. The energy absorption per atom increases with the cluster size and saturates for larger clusters. This tendency is in qualitative agreement with the experimental data on the ejection energies of three different ion charge states, +1, +4 and +7 (fig. 3.11 in ref.~\cite{desy2}). 

We also list our predictions on the ionization and thermalization timescales for electrons within the clusters. These timescales are measured from the start of the exposure at $t=0$ fs. We checked that if the cluster is irradiated with single rectangular VUV pulses of a flux fixed to $I\cdot \Delta t=0.4$ J/cm$^2$, saturation of ionizations from ground states occurs on average after: $t_I \sim 2.5-3 \cdot \Delta t$. For longest pulses ($\Delta t \sim 50$ fs) $t_I \sim \Delta t$ and for shorter pulses ($\Delta t \sim 4$ fs) $t_I \sim 4\Delta t$.

The thermalization timescale is defined as the timescale needed to achieve the LTE within a system. After this time the velocity distribution of electrons approaches the Maxwell-Boltzmann distribution. Within the large clusters only a small fraction of electrons can leave the sample. The remaining electrons are then heated up inside the cluster by the IB process, and they thermalize fast. As the non-equilibrium electron escape dominates the evolution of smaller clusters early in the exposure, the thermalization timescales for small 
clusters will be longer than for the large clusters.   
Thermalization timescales, $t_T$, estimated at this particular integrated radiation flux, $F=I\cdot \Delta t=0.4$, J/cm$^2$ are: i) for small clusters (20 and 70 atoms) $0.8 \cdot \Delta t< t_T < 4 \cdot \Delta t$, and ii) for large clusters (2500 and 90000 atoms) $0.6 \cdot \Delta t < t_T < 2.5 \cdot \Delta t$,
depending on the pulse length. For comparison, in ref.~\cite{ziajab2} the thermalization timescale for the cluster $Xe_{2500}$ at the higher integrated radiation flux of $F=0.6$ J/cm$^2$ was found to be $\geq 0.3\cdot \Delta t$. 

To sum up, we used a microscopic model based on the first principle Boltzmann approach to investigate non-equilibrium dynamics within atomic clusters of various sizes irradiated by single VUV pulses.  This model is computationally efficient for small and large clusters, and includes various processes that are relevant at VUV photon energies. In particular, it uses the enhanced IB rate
obtained using effective atomic potential as proposed in ref.~\cite{santra,santra1} and accounts also for the plasma environment effects: the electron screening \cite{murillo} and the lowering of the interatomic potential barriers (following the ideas of refs.~\cite{jort3,siedschlag,georg}).
Within our model cold electrons released during photoionizations are heated by the IB process. Highly charged ions are created during collisions of energetic quasi-free electrons with ions. Electrostatic interactions among charges, interactions with the external laser field and shielded electron-electron interactions are treated. Recombination processes are neglected due to their low rates expected and short simulation timescales ($\leq 100$ fs) \cite{ziajab2}.

The results obtained demonstrate the different ionization dynamics of small and large clusters. Predictions obtained at various cluster sizes on: (i) the maximal and average ion charge, (ii) ion fractions  within the outer shell, and (iii) average energy absorbed per atom were compared to the experimental data. They were in good agreement with these data. However, a quantitative verification of the outer shell predictions should be performed with an expansion code. 

We have shown that our non-equilibrium Boltzmann solver is a useful tool for studying the evolution of irradiated samples. Several interesting problems may be investigated in future with this code and with an expansion code, e.\ g.\  the evolution of clusters irradiated with XUV and X-rays, the evolution of 
mixed clusters or mechanisms of slowing down the cluster explosion. 

Also, if ever ultrashort FEL pulses are available ($1 \leq \Delta t \leq 10$ fs) \cite{saldin2,saldin3,saldin4,saldin1}, the evolution of the samples irradiated with these pulses may be fully non-equilibrium during the exposure. As the applicability of the semi-equilibrium hydrodynamic models at these pulse lengths is much restricted, we expect that the non-equilibrium Boltzmann solver can be particularly useful for studying the properties of larger samples irradiated with ultrashort FEL pulses.

\acknowledgments
Beata Ziaja is grateful to Cornelia Deiss, Wojciech Rozmus, Robin Santra and Abraham Sz\"oke for illuminating comments. Thomas M\"oller thanks the colleagues from Dresden, especially Jan Michael Rost and Ulf Saalmann, as well as Joshua
Jortner (Tel Aviv) for fruitful discussions. 
This research was supported by the German Bundesministerium f\"ur Bildung und Forschung with grants No.\ 05 KS4 KTC/1, No.\ 05 KS7 KT1 and by the Helmholtz
Gemeinschaft, Impulsfond VH-VI-103. 

\begin{thebibliography}{10}
\expandafter\ifx\csname url\endcsname\relax\def\url#1{\texttt{#1}}\fi

\bibitem{tesla}
\Name{DESY} \REVIEW{{TESLA, the Superconducting Electron-Positron Linear
  Collider with an integrated X-ray Laser Laboratory. Technical Design
  Report.,DESY, ISBN 3-935702-00-0} }{5}{2001}{150}.

\bibitem{desy2006}
\Name{DESY} \REVIEW{{Technical Design Report of the European XFEL, DESY, ISBN
  3-935702-17-5} }{5}{2006}{7}.

\bibitem{slac}
\Name{LCLS} \REVIEW{{LCLS: The First Experiments., SSRL, SLAC, Stanford, USA}
  }{}{2000}{}.

\bibitem{jap}
\Name{Shintake T. \and Team S.} \REVIEW{{Proceedings of FEL 2006, BESSY,
  Berlin, Germany} }{}{2006}{33}.

\bibitem{xfelinfo2007}
\Name{DESY} \REVIEW{DESY, {http://xfelinfo.desy.de} }{5}{2007}{18}.

\bibitem{l1}
\Name{Neutze R., Wouts R., van~der Spoel D., Weckert E. \and Hajdu J.}
  \REVIEW{Nature }{406}{2000}{752}.

\bibitem{miao}
\Name{Miao J., Hodgson K. \and Sayre D.} \REVIEW{Proc. Natl. Acad. Sci.
  }{98}{2001}{6641}.

\bibitem{gyula1}
\Name{Jurek Z., Oszl\'anyi G. \and Faigel G.} \REVIEW{Europhys. Lett.
  }{65}{2004}{491}.

\bibitem{plasma4}
\Name{{S. P. Hau-Riege et al.}} \REVIEW{Phys. Rev. E }{71}{2005}{061919}.

\bibitem{liver1en}
\Name{{S. P. Hau-Riege et al.}} \REVIEW{Phys. Rev. Lett. }{98}{2007}{198302}.

\bibitem{chapman}
\Name{{H. Chapman et al.}} \REVIEW{Nature Physics }{2}{2006}{839}.

\bibitem{ziajab2}
\Name{Ziaja B., Wabnitz H., Weckert E. \and M\"oller T.} \REVIEW{{arXiv:
  physics/0711.3725, submitted for publication} }{}{2007}{}.

\bibitem{desy}
\Name{{H. Wabnitz et al.}} \REVIEW{Nature }{420}{2002}{482}.

\bibitem{desy2}
\Name{Wabnitz H.} \REVIEW{Doctoral Thesis}{DESY-THESIS-2003-026}{2003}{}.

\bibitem{desy3}
\Name{{T. Laarmann et al.}} \REVIEW{Phys. Rev. Lett. }{95}{2005}{063402}.

\bibitem{ziajab}
\Name{Ziaja B., de~Castro A. R.~B., Weckert E. \and M\"oller T.} \REVIEW{Eur.
  Phys. J. D }{40}{2006}{465}.

\bibitem{santra}
\Name{Santra R. \and Greene C.~H.} \REVIEW{Phys. Rev. Lett.
  }{91}{2003}{233401}.

\bibitem{santra1}
\Name{Walters Z.~B., Santra R. \and Greene C.~H.} \REVIEW{Phys. Rev. A
  }{74}{2006}{043204}.

\bibitem{gyula}
\Name{Jurek Z., Faigel G. \and Tegze M.} \REVIEW{Eur. Phys. D }{29}{2004}{217}.

\bibitem{plasma}
\Name{Hau-Riege S.~P., London R.~A. \and Sz\"oke A.} \REVIEW{Phys. Rev. E
  }{69}{2004}{051906}.

\bibitem{siedschlag}
\Name{Siedschlag C. \and Rost J.} \REVIEW{Phys. Rev. Lett. }{93}{2004}{043402}.

\bibitem{rost1}
\Name{Saalmann U., Siedschlag C. \and Rost J.~M.} \REVIEW{J. Phys. B
  }{39}{2006}{R39}.

\bibitem{water}
\Name{Bergh M., Timneanu N. \and van~der Spoel D.} \REVIEW{Phys. Rev. E
  }{70}{2004}{051904}.

\bibitem{desy7}
\Name{{T. Laarmann et al.}} \REVIEW{Phys. Rev. Lett. }{92}{2004}{143401}.

\bibitem{murillo}
\Name{Murillo M.~S. \and Weisheit J.~C.} \REVIEW{Physics Reports
  }{302}{1998}{1}.

\bibitem{jort3}
\Name{Last I. \and Jortner J.} \REVIEW{Physical Review A }{62}{2000}{013201}.

\bibitem{georg}
\Name{Georgescu I., Saalmann U. \and Rost J.-M.} \REVIEW{Phys. Rev. A
  }{76}{2007}{043203}.

\bibitem{saldin2}
\Name{Saldin E.~L., Schneidmiller E.~A. \and Yurkov M.~V.} \REVIEW{Opt. Commun.
  }{212}{2002}{377}.

\bibitem{saldin3}
\Name{Saldin E.~L., Schneidmiller E.~A. \and Yurkov M.~V.} \REVIEW{Opt. Commun.
  }{237}{2004}{153}.

\bibitem{saldin4}
\Name{Saldin E.~L., Schneidmiller E.~A. \and Yurkov M.~V.} \REVIEW{Opt. Commun.
  }{239}{2004}{161}.

\bibitem{saldin1}
\Name{Saldin E.~L., Schneidmiller E.~A. \and Yurkov M.~V.} \REVIEW{{Phys. Rev.
  ST Accel. Beams} }{9}{2006}{050702}.

\end{thebibliography}

\end{document}